\documentclass[pre,twocolumn,showpacs,showkeys,superscriptaddress,preprintnumbers,floatfix]{revtex4-1}

\usepackage{etex}
\usepackage{ifpdf}
\usepackage{hyperref}
\usepackage{dcolumn}
\usepackage{url}
\usepackage{amsmath}
\usepackage{amscd}
\usepackage{amsfonts}
\usepackage{amssymb}
\usepackage{amsthm}
\usepackage{xfrac}
\usepackage{braket}
\usepackage{bm}   
\usepackage{bbm}
\usepackage{verbatim}
\usepackage{mathtools}
\usepackage{stmaryrd}
\usepackage{xcolor}
\usepackage{setspace}
\usepackage{tikz}
\usetikzlibrary{arrows}
\usetikzlibrary{automata}
\usetikzlibrary{calc}
\usetikzlibrary{decorations.pathreplacing}
\usetikzlibrary{patterns}
\usetikzlibrary{positioning}
\usetikzlibrary{plotmarks}
\usetikzlibrary{shapes}

\usepackage{pgfplots}
\pgfplotsset{compat=newest}

\tikzstyle{vaucanson}=[
  node distance=2.5cm,
  bend angle=15,
  auto,
  every loop/.style={},
  every edge/.style={->,draw=black,line width=1.2,>=latex,shorten <=1pt,shorten >=1pt},
  every state/.style={draw=black,line width=2,font=\large},
  loop right/.style={right,out=22,in=-22,loop},
  loop above/.style={above,out=112,in=68,loop},
  loop left/.style={left,out=202,in=158,loop},
  loop below/.style={below,out=292,in=248,loop},
  loop above right/.style={right,out=67,in=23,loop},
  loop above left/.style={left,out=157,in=113,loop},
  loop below left/.style={left,out=247,in=203,loop},
  loop below right/.style={right,out=337,in=293,loop},
]

\theoremstyle{plain}    
\theoremstyle{plain}    
\theoremstyle{plain}    
\theoremstyle{plain}    
\theoremstyle{plain}    
\theoremstyle{plain}    
\theoremstyle{plain}    
\theoremstyle{plain}    
\theoremstyle{plain}    
\theoremstyle{plain}    
\theoremstyle{plain}    
\theoremstyle{plain}    
\theoremstyle{plain}    
\theoremstyle{plain}    
\theoremstyle{plain}    
\theoremstyle{plain}    
\theoremstyle{plain}

\newcommand{\MeasAlphabet}  {\mathcal{A}}
\newcommand{\MeasSymbol}   { {X} }
\newcommand{\meassymbol}   { {x} }

\newcommand{\CausalState}   { \mathcal{S} }

\newcommand{\causalstate}   { \sigma }
\newcommand{\CausalStateSet}    { \boldsymbol{\CausalState} }
\newcommand{\AlternateState}    { \mathcal{R} }

\newcommand{\AlternateStateSet} { \boldsymbol{\AlternateState} }

\newcommand{\Prob}      {\Pr} 

\newcommand{\EE}        {{\bf E}}

\newcommand{\ProcessAlphabet}   {\MeasAlphabet}

\newcommand{\forward}{+}
\newcommand{\reverse}{-}
\newcommand{\forwardreverse}{\pm} 

\newcommand{\FutureCausalState} { {\CausalState}^{\forward} }

\newcommand{\PastCausalState}   { {\CausalState}^{\reverse} }

\newcommand{\lastindex}[2]{
  \edef\tempa{0}
  \edef\tempb{#2}
  \ifx\tempa\tempb
    \edef\tempc{#1}
  \else
    \edef\tempa{0}
    \edef\tempb{#1}
    \ifx\tempa\tempb
      \edef\tempc{#2}
    \else
      \edef\tempc{#1+#2}
    \fi
  \fi
  \tempc
}

\newcommand{\CSjoint}[1][,]{
   \edef\tempa{:}
   \edef\tempb{#1}
   \ifx\tempa\tempb
      \ensuremath{\FutureCausalState\!#1\PastCausalState}
   \else
      \ensuremath{\FutureCausalState#1\PastCausalState}
   \fi
}

\newif\ifpm
\edef\tempa{\forwardreverse}
\edef\tempb{\pm}
\ifx\tempa\tempb
   \pmfalse
\else
   \pmtrue
\fi

\colorlet {R_color}    {blue}
\colorlet {k_color}    {black!30!green}

\def\clap#1{\hbox to 0pt{\hss#1\hss}}

\begin{document}

\title{Time cells might be optimized for predictive capacity, \\ not redundancy reduction or memory capacity}

\author{Alexander Hsu}

\author{Sarah E. Marzen}
\email{smarzen@cmc.edu}
\affiliation{W. M. Keck Science Department, Claremont, CA 91711}

\date{\today}
\bibliographystyle{unsrt}

\begin{abstract}
Recently, researchers have found time cells in the hippocampus that appear to contain information about the timing of past events.  Some researchers have argued that  time cells are taking a Laplace transform of their input in order to reconstruct the past stimulus.  We argue that stimulus prediction, not stimulus reconstruction or redundancy reduction, is in better agreement with observed responses of time cells.  In the process, we introduce new analyses of nonlinear, continuous-time reservoirs that model these time cells.
\end{abstract}


%
\preprint{arxiv.org:1702.08565 [physics.gen-ph]}

\maketitle


\setstretch{1.1}

\newcommand{\Abet}{\ProcessAlphabet}
\newcommand{\MS}{\MeasSymbol}
\newcommand{\ms}{\meassymbol}
\newcommand{\SSet}{\CausalStateSet}
\newcommand{\St}{\CausalState}
\newcommand{\st}{\causalstate}
\newcommand{\MxSt}{\AlternateState}
\newcommand{\MxSSet}{\AlternateStateSet}
\newcommand{\mxst}{\mu}
\newcommand{\mxstt}[1]{\mu_{#1}}
\newcommand{\StartMS}{\bra{\delta_\pi}}
\newcommand{\Ipred}{\EE}
\newcommand{\ISI} { \xi }

\newcommand{\ECT}{\widehat{\EE}}
\newcommand{\CCT}{\widehat{C}_\mu}

\newcommand{\gen}{g}
\newcommand{\FeatAlphabet}{\mathcal{F}}


\vspace{0.2in}
\section{Introduction}
Recent experiments have revealed the presence of so-called time cells in the hippocampus, which seem to fire to signal the timing of a certain event \cite{eichenbaum_2014}.  Time cells fire even when location information or behavioral information is constant \cite{kraus2013hippocampal}, and are thought to support episodic memory-- memory of what, where, and when an event was experienced \cite{eichenbaum_2014}.

Ref. \cite{howard2014} offers a novel explanation of time cells, which applies to not just temporal signals, but also spatial signals and others: they claim that time cells are computing a Laplace transform of the input, and that the past input is linearly reconstructed from discrete samples of this Laplace transform.  These model time cells are therefore linear continuous-time reservoirs, or linear echo state networks \cite{hermans2010memory, lukovsevivcius2009reservoir, jaeger2001echo}, which can simulate, predict, and remember limited types of input.  Their nonlinear counterparts can simulate any type of input with enough nodes (neurons) \cite{grigoryeva_2018}.

Implicit in several descriptions of time cells \cite{macdonald_2011,eichenbaum_2014,jazayeri_2015} is that the goal of these cells is to reconstruct the past stimulus.  This certainly seems like a worthwhile goal for an organism.  However, some classic work suggests that neurons try to ``efficiently code'' their stimulus minimize redundancy \cite{barlow_1961}, and some recent works have suggested that the goal of some biological subsystems is to predict the future, e.g. as in Refs. \cite{palmer2015predictive, glimcher2011understanding}.  These goals might all sound similar, and to some extent they are-- one needs memory to predict, for example.  But it is also possible to have infinite memory and no predictive power \cite{marzen_2017}.  Here, we compare predictions of each of these normative principles to ascertain which are consistent with observed time cell properties.  To do so, we extended the results and the methodologies of Ref. \cite{marzen_2017} to the case of some nonlinear and all linear continuous-time reservoirs, thus extending the work of Ref. \cite{hermans2010memory}.


Only maximization of predictive power of time cells when stimulated with naturalistic stimuli yields neuronal timescales that behave near to what is seen in experiment \cite{howard2014}, suggesting that prediction-- not reconstruction or redundancy reduction-- may be key to understanding the properties of time cells.  This conclusion assumes both that natural video's autocorrelation function does not have significant oscillatory components and that the brain also has ``readout neurons'' that simply communicate information about only the present stimulus.  Prediction has already proven key for understanding other aspects of neural processing \cite{palmer2015predictive, glimcher2011understanding}.

The paper starts by describing our setup, in which we specialize to a stationary stimulus and the normative principles listed above.  We then describe the time scales of model neurons that minimize redundancy, maximize memory, or maximize prediction for both simple and more naturalistic stimuli, and show that only maximization of predictive ability might match experiment.


\section{Setup}

The organism is exposed to a continuously varying temporal signal $\overleftrightarrow{x}$, whose value at time $t$ is $x_t$.  For ease, we assume that the stimulus is a scalar with zero mean $\langle x_t\rangle = 0$ and unit variance $\langle x_t^2\rangle =1$.  This temporal signal is a realization of an ergodic stationary stochastic process with random variable $\overleftrightarrow{X}$ symbolizing the whole signal and $\overrightarrow{X}_t^T$ symbolizing the trajectory that starts at $t$ and ends at $t+T$.  Stationarity implies that $\Prob(\overrightarrow{X}_t^T)$ is independent of $t$, and ergodicity implies that different realizations have identical statistics.

We assume that the autocorrelation function of the input signal can be written as
\begin{equation}
R(t) = \int_0^{\infty} F(\lambda) e^{-\lambda |t|} d\lambda.
\end{equation}
All autocorrelation functions can be written in this form if one extends the integral to exist over the complex plane.  In this manuscript, we study exponential autocorrelation functions and oscillatory and exponentially decaying autocorrelation functions.  In the latter case, we can use the formulae developed later by allowing $F(\lambda)$ to have support on imaginary numbers with negative real parts.

Three types of input are studied: a particle moving according to an overdamped Langevin equation; a particle moving according to an underdamped Langevin equation; and a particle whose position has statistics similar to that of natural video.  In the first case, we approximate the autocorrelation function $R(t)$ as a single exponential,
\begin{equation}
    R(t) = e^{-\lambda |t|}.
\end{equation}
In the second case, we approximate the autocorrelation function $R(t)$ as a decaying exponential multiplied by an oscillatory function,
\begin{equation}
    R(t) = e^{-\lambda |t|} \cos(\omega t).
\end{equation}
In the second case, we turn to Ref. \cite{dong_atick_1995}, in which it was found that the power spectrum of natural video is roughly $\frac{1}{|\omega|^\alpha}$ for $\alpha$ between 1 and 2, resulting in power law autocorrelation functions with exponents between 1 and 2. In order to study model time cells under naturalistic conditions, we consider autocorrelation functions of the form 
\begin{equation}
    R(t)=\frac{1}{1+|t|^\alpha}.
\end{equation}
Some of our results will hold more generally than for just these three conditions.

The organism is presumed to have model time cells whose activity changes as a function of sensory signal-- so-called time cells.  These neurons might have their response properties tuned based on one of many normative principles that we discuss below.

Finally, for what follows, we need to define the entropy of a random variable $Y$ with realizations $y\in\mathcal{Y}$, and the mutual information of a random variable $Y$ and another random variable $Z$ with realizations $z\in\mathcal{Z}$.  The entropy $H[Y]$ is given by $-\sum_y p(y)\log p(y)$, and the mutual information $I[Y;Z]$ is given by $\sum_{y,z} p(y,z)\log\frac{p(y,z)}{p(y)p(z)}$.  The entropy is the uncertainty, while the mutual information captures the reduction in uncertainty we achieve by knowing one of the variables \cite{Cove91a}.  There are, of course, operational meanings to entropy and mutual information via Shannon's theorems, but we do not need these theorems for what follows.

\subsection{Model of time cells}

We are interested in two types of model time cells.  The first type of time cell merely remembers what it saw a time $s$ in the past.  The second type of time cell follows the formulation of Ref. \cite{howard2014}, as it computes a Laplace transform of the input.  In the main text, we will only consider the second type.  Results for the first type, which are qualitatively similar, are in the appendix.

The activity of time cell $f(t)$ with neuronal forgetting rate $s$ (an inverse neuronal time scale) at time $t$ is
\begin{equation}
f(t) = \int_0^{\infty} e^{-st'} x(t-t') dt',
\end{equation}
which can be achieved via a leaky integrator,
\begin{equation}
\frac{df}{dt} = -sf + x.
\end{equation}
This is a Laplace transform but sampled only at some values of $s$.  The stimulus $\overleftarrow{x}_t$ can be inferred by an approximate inverse Laplace transform or (nearly equivalently) by an optimal linear estimate.

We imagine that there are $N$ neurons, and that the $i^{th}$ neuron has a forgetting rate $s_i$.  We order the neurons without loss of generality so that $\{s_i\}_{i=1}^N$ is monotonically increasing.  The neural activity of time cell $i$ at time $t$ is denoted $f_{s_i}(t)$.

Although our setup might seem limited in that these recurrent networks are ``simple''-- that is, there are only self-loops and no connections between neurons-- simple linear recurrent networks are just as powerful as the more complex linear recurrent networks with connections between different neurons.  This fact comes from Ref. \cite{marzen_2017} and the formulae derived in the subsection below, and is only true when recurrent networks are linear.

One might expect a qualitatively different story when the activities are nonlinear functions of past input, but in the appendix we show that linearity is desirable for maximal predictive capacity.  Still, a full understanding of nonlinear reservoirs will be the subject of future work.

\subsection{Variety of normative principles}

There are at least four normative principles that could explain the properties of time cells: minimization of redundancy \cite{barlow_1961,howard_2018} between neighboring time cells; maximization of memory capacity, which is a metric for how well one can reconstruct the past stimulus from the present neuronal response \cite{jaeger2001short, white2004short, boedecker2012information, farkavs2016computational, baranvcok2014memory}; maximization of the joint entropy of all neuronal responses, as derived from the efficient coding hypothesis \cite{barlow_1961}, which is sometimes rephrased as redundancy reduction; and maximization of predictive capacity, which is a metric for how well one can predict the future stimulus from the present neuronal response \cite{marzen_2017}.

Each of these normative principles is quantified as follows.  Redundancy, as is typical, is deemed to be the mutual information between the output of two neurons.
We extend the definition of discrete-time memory capacity \cite{marzen_2017} and predictive capacity \cite{marzen_2017} to continuous-time via
\begin{equation}
MC = \int_{-\infty}^0 m(\tau) d\tau,~PC = \int_0^{\infty} m(\tau) d\tau
\end{equation}
where the memory function $m(\tau)$ is the squared correlation coefficient between the optimal linear estimate of $x(t+\tau)$ using $f(t)$, which one can show is:
\begin{equation}
m(\tau) = \langle f(t) x(t+\tau)\rangle_t^{\top} \langle f(t) f(t)^{\top}\rangle_t^{-1} \langle f(t) x(t+\tau)\rangle_t.
\end{equation}
We have assumed that the input is zero-mean and of unit variance.
Although it seems unlikely that an organism is interested in arbitrarily long pasts, the infinite limit provides good intuition for the more biophysically reasonable, finite-time case.
In the appendix, we provide a derivation of the following closed-form expression for $MC$:
\begin{equation}
MC = 1^{\top} \left( C^{-1} \odot D_{MC} \right) 1
\end{equation}
where
\begin{equation}
C_{ij} = \int_0^{\infty} F(\lambda) \frac{2\lambda+s_i+s_j}{(\lambda+s_i)(\lambda+s_j)(s_i+s_j)} d\lambda
\end{equation}
and
\begin{widetext}
\begin{eqnarray}
\left(D_{MC}\right)_{i,j}=\int_{\lambda=0}^{\infty}\int_{\lambda'=0}^{\infty}F(\lambda)F(\lambda')\left(\frac{1}{\left(\lambda^{2}-s_{i}^{2}\right)\left(\lambda'^{2}-s_{j}^{2}\right)}\left[\frac{4\lambda\lambda'}{s_{i}+s_{j}}-\frac{2\lambda(\lambda'+s_{j})}{s_{i}+\lambda'}-\frac{2\lambda'(\lambda+s_{i})}{\lambda+s_{j}}+\frac{(\lambda+s_{i})(\lambda'+s_{j})}{\lambda+\lambda'}\right]\right)d\lambda' .
\label{eq:DMC}
\end{eqnarray}
\end{widetext}
Furthermore, also in the appendix, we show that
\begin{equation}
PC = 1^{\top} \left(C^{-1} \odot D_{PC}\right) 1
\end{equation}
where $C_{ij}$ is as before and
\begin{equation}
(D_{PC})_{ij} = \int_0^{\infty} \int_0^{\infty}\frac{F(\lambda) F(\lambda')}{(\lambda+\lambda')(\lambda+s_i)(\lambda'+s_j)} d\lambda d\lambda'.
\label{eq:DPC}
\end{equation}
Note that these formulae also allow for complex $\lambda$, if the integrals or sums are appropriately extended.  This is quite useful for oscillatory input.

These formulae are complicated, but they lead to two main points.  First, the \emph{only} relevant environmental statistics for $MC$ and $PC$ are the autocorrelation function of the input.  This is true also for the discrete-time case.  Hence, stimulating time cells with real natural video will, in theory, yield the same $MC$ or $PC$ as usage of the above formulae with the autocorrelation function of naturalistic input.  Second, the formulae above may yield more accurate calculations of $MC$ or $PC$, as we have traded difficulties associated with too little data for difficulties of accurate numerical integration and matrix inversion.  Which of these difficulties is more pressing will depend on one's application.

We could also consider combinations of the above normative principles.  For instance, one might try to maximize predictive power while minimizing memory, as in Refs. \cite{still2010optimal, palmer2015predictive, marzen2016predictive, chalk2018toward}.  We discuss this possibility later, but shy away from doing a combination of optimization principles in this paper because it is likely possible to achieve almost any desired optimal neural forgetting rate by appropriate choice of Lagrange multipliers.

\section{Results}

In what follows, we derive the optimized neuronal time scales for each of the normative principles for three types of input: a particle moving according to an overdamped Langevin equation; a particle moving according to an underdamped Langevin equation; and a particle whose position has an autocorrelation function like that of natural videos.

For our linear time cells, as stated earlier, only the autocorrelation function of the input affects predictive capacity and memory capacity.  (See the appendix.)  This is a theoretical conclusion that greatly simplifies any effort to find optimal neuronal forgetting rates, as we only need to estimate the autocorrelation function of natural input and input such autocorrelation functions into the formulae given earlier and in our appendices.  In our two toy examples, the autocorrelation function takes the form of a single exponential (overdamped) and an oscillatory decaying exponential (underdamped) with Gaussian statistics, so that again, only the autocorrelation function determines memory capacity, predictive capacity, and also redundancy.  Because the autocorrelation function uniquely determines memory and predictive capacity, the memory and predictive capacities given here for naturalistic input are the same as if we had simulated our model time cells being stimulated with natural video.


\subsection{Redundancy equalization and minimization}

It seems desirable to reduce redundancy between neurons \cite{barlow_1961}.  Two simple examples will illustrate that redundancy minimization do not typically yield logarithmic scaling, as anticipated by Ref. \cite{howard_2018}.  Suppose that $x(t)$ is a Gaussian process, which is necessarily true for outputs of overdamped and underdamped Langevin equations.  A straightforward calculation then gives
\begin{eqnarray}
I[f_{s_i}(t);f_{s_{i+1}}(t)] &=& \log\sqrt{\frac{1}{1-\rho^2}},
\end{eqnarray}
where $\rho^2 = \frac{\langle f_{s_i}(t) f_{s_{i+1}}(t) \rangle^2}{\langle f_{s_i}(t)^2\rangle \langle f_{s_{i+1}}(t)^2\rangle}$ is the correlation coefficient for zero-mean processes.  Some straightforward algebra reveals that
\begin{widetext}
\begin{equation}
    \rho^2 = \frac{\left(\int_0^{\infty}\int_0^{\infty} e^{-s_i t}e^{-s_j t'} R(t-t') dt dt'\right)^2}{\left(\int_0^{\infty} \int_0^{\infty} e^{-s_i (t+t')} R(t-t') dt dt'\right)\left( \int_0^{\infty} \int_0^{\infty} e^{-s_j (t+t')} R(t-t') dt dt'\right)}.
\end{equation}
\end{widetext}
This mutual information is a typical measure of redundancy \cite{howard_2018}.  Redundancy is minimized when the correlation coefficient is minimized.

In the overdamped case, a straightforward calculation gives, for $s_{i+1}=\Delta_i s_i$,
\begin{eqnarray}
\rho^2 &=& \frac{\Delta_i}{(1+\Delta_i)^2} \frac{(2\lambda+s_i+\Delta_i s_i)^2}{(\lambda+s_i)(\lambda+\Delta_i s_i)}.
\end{eqnarray}
To equalize redundancy between two successive sets of neurons, we must set $\rho^2$ to be constant, which cannot be accomplished for this type of input. Some algebra reveals that equalized redundancy implies negative neuronal timescales, a biophysical impossibility.  In fact, redundancy equalization is either unachievable or does not seem to imply logarithmic scaling \emph{unless} the input has exactly power-law autocorrelation as was found in Ref. \cite{howard_2018}, based on calculations not shown here.

To minimize redundancy when the input moves according to an overdamped Langevin equation, we must make forgetting rates $s_i,~s_{i+1}$ as big as possible, while making $\Delta_i$ as large as possible as well. No matter the input, we tend to find that neurons should all forget past stimulus information as quickly as possible.

Typically, e.g. when performing independent components analysis \cite{brown2001independent}, one finds that redundancy is reduced when different neurons pick up on orthogonal aspects of the stimulus.  With this Laplace transform model of time cells, such decoupling is not possible, and reducing redundancy requires sending at least one of the neuronal forgetting rates to infinity.
In particular, to minimize redundancy when the input moves according to an underdamped Langevin equation, or when the input's position has naturalistic statistics, some numerical experiments suggest that we must make forgetting rates as dissimilar as possible, i.e. $s_i\rightarrow 0,~s_{i+1}\rightarrow\infty$.  This intuitively makes some sense: to reduce correlation between neurons, we should make their responses as dissimilar as possible.

In all cases, to minimize redundancy, we desire to set at least one of the forgetting rates to be infinite, so as to decouple the neurons as much as possible.

\subsection{Efficient coding}

Usually the efficient coding hypothesis \cite{barlow_1961} is phrased as follows: we desire the channel $p(y|x)$ that maximizes mutual information subject to a capacity constraint, $p^*(y|x) := \arg\max_{p(y|x):I[X;Y]\leq C} I[X;Y]$.  This, alone, is underdetermined, and so we also impose another constraint: that $p(y|x)$ be a deterministic mapping, so that $I[X;Y] = H[Y] - H[Y|X] = H[Y]$.  Hence, we are searching for neural responses that maximize the joint entropy, $H[\{f_{s_i}(t)\}_{i=1}^N]$.

As it turns out, this objective function is directly related to the redundancy objective function described in the previous subsection.  Repeatedly using the information theory identity $H[X;Y] = H[X] + H[Y|X]$ yields
\begin{eqnarray}
H[\{f_{s_i}(t)\}_{i=1}^N] &=& H[f_{s_1}(t)] + H[f_{s_2}(t)|f_{s_1}(t)] + \ldots \nonumber \\
&& + H[f_{s_N}(t)|f_{s_1}(t),\ldots ,f_{s_{N-1}}(t)].
\end{eqnarray}
An approximate Markovianity property holds, in that $f_{s}(t)$ is more strongly correlated with the far past when $s$ is smaller, and so $H[f_{s_j}(t)|f_{s_1}(t),\ldots ,f_{s_{j-1}}(t)]$ is approximately $H[f_{s_j}(t)|f_{s_{j-1}}(t)]$.  (This conditional entropy is an upper bound, achievable in the limit that $s_j-s_{j-1} \rightarrow\infty$, but which holds approximately when $s_j-s_{j-1}$ is very large.)  Then, maximizing the joint entropy is approximately equivalent to maximizing $H[f_{s_j}(t)|f_{s_{j-1}}(t)]$, which is equivalent to $H[f_{s_j}(t)] - I[f_{s_j}(t);f_{s_{j-1}}(t)]$.
To the extent that $H[f_{s_j}(t)]$ is roughly constant because $s_j$ is so large that its statistics are governed mostly by the present input, we are left with a minimization of $I[f_{s_j};f_{s_{j-1}}(t)]$-- exactly the objective function of the previous section.  Hence, the results about redundancy reduction hold for the efficient coding hypothesis, even though the objective functions are not exactly the same.

\subsection{Recollecting the past}

There are a number of ways to measure memory, but we focus on the simplest measure (memory capacity $MC$) that was invented to calibrate the performance of reservoir computers \cite{marzen_2017}.  

When the input has a single dominant time-scale as in the overdamped Langevin equation, a glance at the expression for $MC$ earlier suggests that $MC$ will be maximized when the neuronal timescale is exactly matched to the input's timescale.  However, this is not the case.  See the appendix.  For input signals that do not have a significant oscillatory component, optimizing memory capacity means sending \emph{all} forgetting rates to $0$, so that at the limit, neurons are essentially estimates of the mean input symbol, even when the mean is zero.  Such input includes both the overdamped Langevin equation and the naturalistic signals considered in this paper.  A sketch of the argument is in the appendix.

\begin{figure}
\centering
\includegraphics[width=0.5\textwidth]{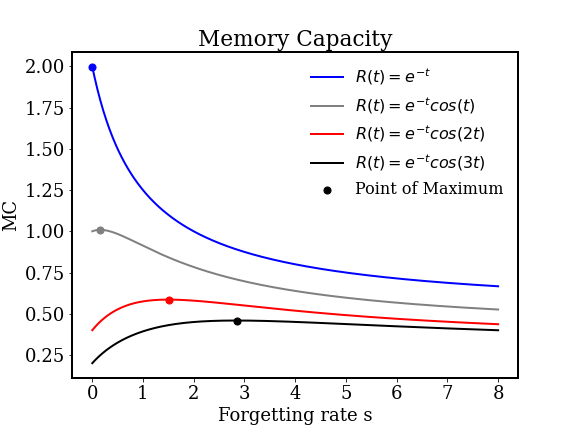}
\caption{A plot of memory capacity $MC$ as a function of neuronal forgetting rate $s$ for a single model time cell (a one-node linear reservoir) for some example autocorrelation functions.  For inputs whose autocorrelation functions may be written as the sum (or integral) of exponentials, $MC$ is maximized when the forgetting rate is $0$ if all of the exponentials have sufficiently small oscillatory components.  See series expansion in the appendix.}
\label{fig:MC}
\end{figure}

Our finding here is similar to what was found for discrete-time reservoir computers \cite{marzen_2017}. An example is shown in Fig. \ref{fig:MC}, where we examine the behavior of memory capacity for some examples of overdamped and underdamped systems.

\begin{figure}
\centering
\includegraphics[width=0.5\textwidth]{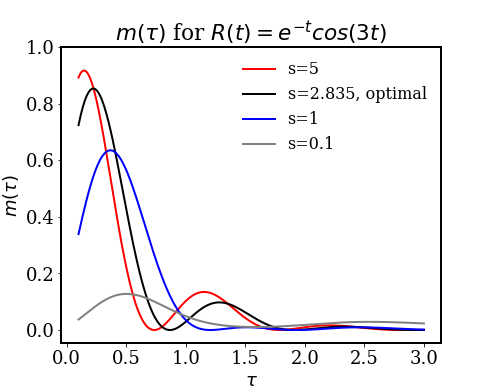}
\caption{A plot of the memory function $m(\tau)$, i.e. the squared correlation coefficient between network state and input a time $\tau$ in the past, for the autocorrelation function $R(t)=e^{-t}\cos(3t)$ and neuronal forgetting rates shown in the legend.  Recall that $MC = \int_0^{\infty} m(\tau) d\tau$.  By appropriately setting the neuronal forgetting rate, you can acquire information about both recent data and data farther in the past with some periodicity.}
\label{fig:mtau_MC}
\end{figure}

When the input has significant oscillatory components, then our argument for setting forgetting rates to $0$ does not hold.  For example, when the input moves according to an underdamped Langevin equation, $MC$ is maximized at a nonzero $s$ .  As the frequency increases in the underdamped system, we find that the optimal forgetting rate generally increases as well (when $R(t)=e^{t}\cos{(\omega t)}$, the optimal $s$ for MC as a function of $\omega$ is approximately piecewise linear). Examining the values of $m(\tau)$ directly, this can perhaps be explained by the fact that remembering recent values very accurately is helpful in remembering the values in the period before.  See Fig. \ref{fig:mtau_MC}.

\begin{figure}
\centering
\includegraphics[width=0.5\textwidth]{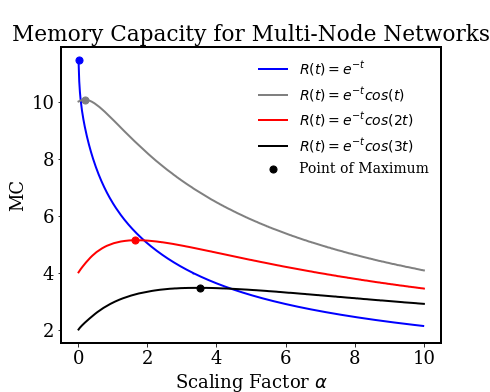}
\caption{Memory capacity as a function of scaling factor $\alpha$ for a network made up of neurons with forgetting rates $s_i = \alpha \frac{i}{10}$ for $i=1,...,10$, for various autocorrelation functions of the input.  When oscillations are of high enough frequency, the optimal $\alpha$ for maximal $MC$ is nonzero.}
\label{fig:multinodeMC}
\end{figure}

The case of multiple neurons seems qualitatively similar to that of the single node case when examining scaling properties. We demonstrate this by examining the case of ten neurons spaced equally between $0$ and $1$, scaling all of them by a factor $\alpha$ and examining the values of $MC$ for networks generated in this manner. See \ref{fig:multinodeMC}. We still find that having a higher frequency component in the underdamped system causes optimal forgetting rates that are greater than zero, unlike the overdamped case.

In conclusion, if an input has significant oscillatory components, then maximizing memory capacity $MC$ may lead to nonzero forgetting rates.  But if the input's autocorrelation function seems to be the sum of decaying exponentials rather than a sum of oscillating and decaying exponentials-- as seems to be true for naturalistic video \cite{dong_atick_1995}-- then a series expansion in the appendix and numerical experiments presented here all suggest that maximizing memory capacity will yield forgetting rates that are as close to zero as possible.

\subsection{Predicting the future}

Finally, we might expect neurons to maximize something like a predictive capacity $PC$, as described earlier.  As we detail in the appendix, perhaps surprisingly, linear recurrent neurons can beat nonlinear recurrent neurons at predicting input.  As such, prediction already in part explains why time cells might want to perform an approximate Laplace transform.

Perhaps not surprisingly, $PC$ is often optimized by setting $s\to\infty$, so that the current neuron acts best to only remember what it has just seen.  This corresponds to the fact that the present signal usually has more information about future signals than past signals.  To illustrate this phenomenon, we consider the impinging process to have an autocorrelation function of $R(t) = \frac{1}{2} e^{-\lambda_1|t|} + \frac{1}{2} e^{-\lambda_2 |t|}$, and ask for the optimal forgetting rate of a single time cell $s$.  There is a considerable region of values $\lambda_1$ and $\lambda_2$ for which this optimal forgetting rate is infinite. This corresponds to having a time cell that simply reads out the current value in the time series and has no memory.

This is not surprising from the perspective of understanding nearly Markovian signals.  Recent stimuli convey more information than past ones, and so to predict optimally, one desires information about the most recent stimulus.  But from another perspective, this is quite surprising.  Earlier results in the static case \cite{whittington2017approximation} have shown that when predictive coding-- minimization of error in predicting the stimulus-- is used to optimize neuronal response properties, nontrivial neuronal weights without needing a time cell that has no recurrent connections.  The key to the differences, in our opinion, are based in differences in setup.  Rather than a supervised learning setting in which neuronal weights are tuned to send an input to a prescribed output, we consider a setting in which there are no weights \emph{between} model time cells (based on Ref. \cite{howard_2014}) and in which there is a learned mapping from infinite past inputs to a future input.  The recurrent weight therefore represents not a connection to other neurons but a statement about about feature extraction: which of the past inputs are most informative about the future input?  And for many input time series, the most informative input is the most recent one.

\begin{figure}
\centering
\includegraphics[width=0.1\textwidth]{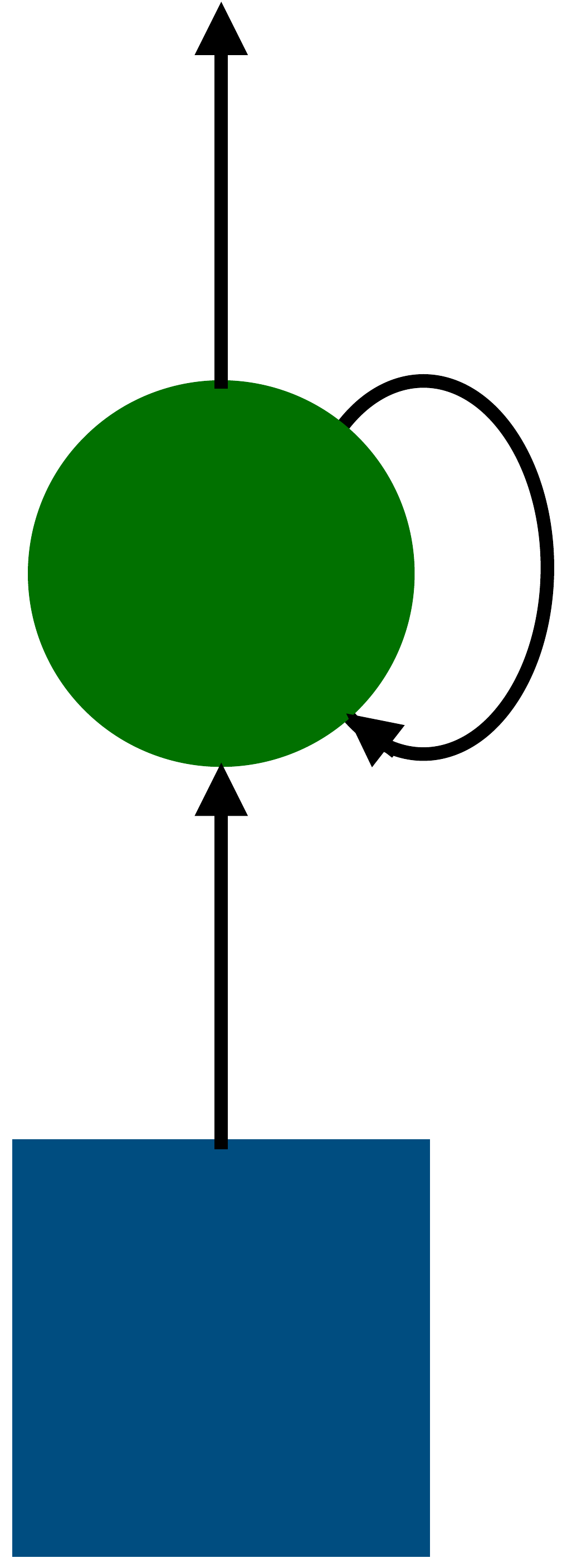}
\hspace{0.1\textwidth}
\includegraphics[width=0.08\textwidth]{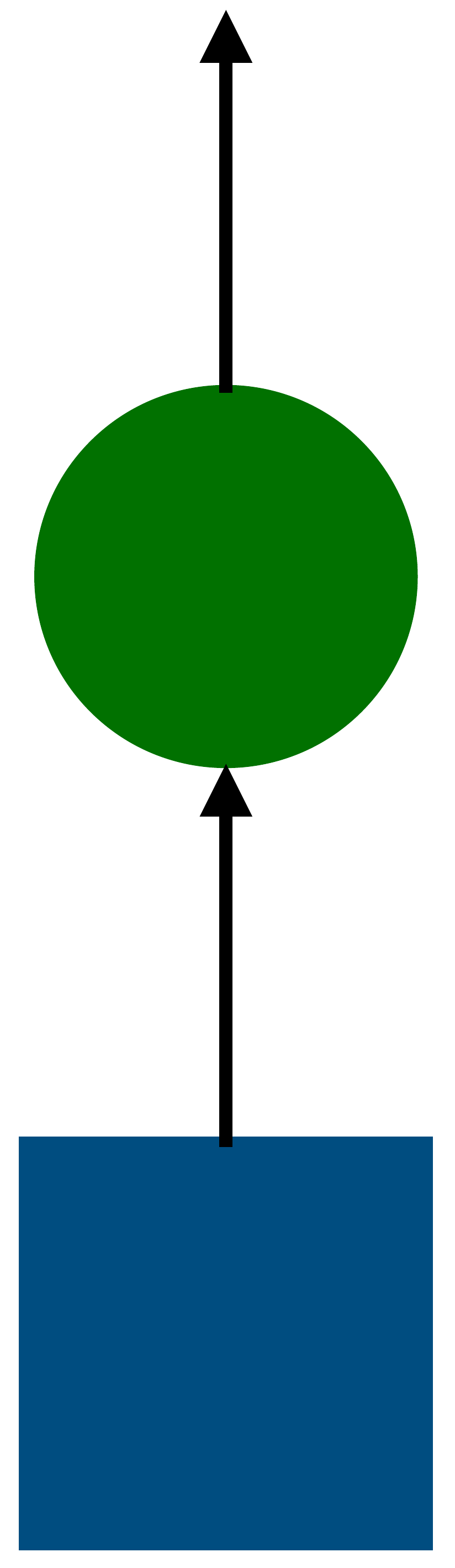}
\caption{A new biological setup that allows for increased predictive capacity.  In both diagrams, the blue square represents the environment and the green circle represents the neuron's activation.  At left, a recurrent neuron representing the current time cell model.  At right, a feedforward neuron that we add to increase predictive capacity, which merely relays current environment information to the downstream region.}
\label{fig:new_setup}
\end{figure}

Thus, we examine time cells with maximal predictive capacity in the presence of an additional cell which explicitly stores the present signal value.  In other words, we imagine the situation shown in Fig. \ref{fig:new_setup}.  Rather than having only cells that take an approximate Laplace transform by implementing a recurrent architecture, we allow for simply one cell to pass through all information about the present input.  This second cell's architecture is entirely feedforward.  It may be biologically relevant that time cells are more predictive when augmented by a single feedforward neuron.  The optimal forgetting rate then is finite and increases with increasing $\lambda_1,~\lambda_2$.  See appendix.

\begin{figure}
\includegraphics[width=0.5\textwidth]{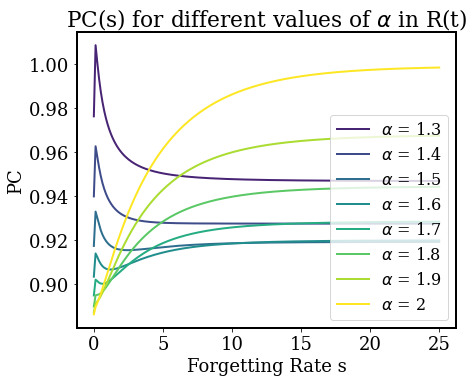}
\caption{Predictive capacity $PC$ as a function of neuronal forgetting rate $s$ for an input with an autocorrelation function $R(t)=\frac{1}{1+|t|^\alpha}$.
$PC(s)$ is shown for various values of $\alpha$.  Intermediate forgetting rates $s$ tend to maximize $PC$ for small enough $\alpha$, though there appears to be  a phase transition in when an intermediate timescale is favored.}
\label{fig:2}
\end{figure} 

As we have just seen, when the input moves according to an overdamped Langevin equation, the optimal neuronal timescale is some nontrivial function of the decay rate.  Similarly, when the input moves according to an underdamped Langevin equation, we continue to see evidence of time-scale matching.  The optimal neuronal timescale is not the oscillatory timescale or the decay timescale, but some non-obvious function of the two.  The authors hope that the equations developed here might aid future efforts to discover this function.

Now we turn our attention to naturalistic signals that have power law autocorrelation functions.  We find that the optimal time constant of the additional time cell is a smoothly varying function of the power law coefficient $\alpha$, assuming that the input is naturalistic. For $\alpha$ between $1$ and $1.789$, $PC$ has a local maximum for a relatively small $s$, i.e. $(s<0.2)$. For $\alpha < 1.56$, this local maximum is the global maximum.  See Fig \ref{fig:2}.
Similarly, Fig. \ref{fig:opts} in the appendix shows the optimal ($PC$-maximizing) time constant of a neuron for the naturalistic stimulus.  Roughly speaking, the optimal time constant of the neuron matches the time constant of the input, a form of time-scale matching not seen in maximization of memory previously.  At $\alpha=1.6$, where the model time cell switches from an optimal intermediate forgetting rate to an optimally maximal forgetting rate.  

\section{Discussion}

Surprisingly, the efficient coding hypothesis, maximization of memory, and redundancy reduction led to the same optimal model time cells when given naturalistic input-- those that remembered the past as well as possible but were relatively useless for understanding the future.  Predictive capacity favored time cells that forgot as much as possible, sans other constraints.  When we included a hand-made neuron that stored the present input, time cells that maximized predictive capacity had time scales tuned to the environment.  We considered the case of higher-dimensional input in the appendix, finding that our main conclusions were unaltered if spatial and temporal components of the spatio-temporal autocorrelation function were separable.

It is worth adding some cautionary words to these sweeping conclusions.  These analyses depend upon exactly what naturalistic input looks like.  We followed Ref. \cite{dong_atick_1995}'s characterization of natural video.  If the autocorrelation function of natural video were later found to be significantly oscillatory, then our results here suggest that maximization of memory capacity could explain observed neuronal forgetting rates.  And some inputs to time cells might easily be oscillatory \cite{mizuseki2009theta}, and for such inputs, maximization of memory capacity would adequately explain nonzero and finite neuronal forgetting rates.

With that aside, to the best of our knowledge now, it seems as though prediction might be closest to the correct normative principle for time cells, as time cells have nonzero forgetting rates \cite{howard2014}.  This may seem strange, as time cells are known for their ability to remember past events.  However, one needs memory for prediction, and so optimizing for prediction does require memory of the ``right'' things \cite{marzen2016predictive}.  For example, remembering what happened $100$ days ago may provide far less useful information as to what will happen tomorrow than remembrance of the previous day's activities.  A more reasonable objective function might be one that balances both memory and prediction, as memory has a coding cost, and prediction is desirable \cite{still2010optimal, palmer2015predictive, marzen2016predictive, chalk2018toward}.  It would be difficult to find the appropriate objective function, however, without fitting to the data, and so we left this potentially thorny issue for future research.


It would also be interesting to see how our conclusions change when the predictive metric is no longer predictive capacity but predictive information \cite{still2010optimal, palmer2015predictive, chalk2018toward}, when memory is explicitly penalized while prediction is valued, when considering nonstationary stimuli, and when considering nonlinear reservoirs for which the Central Limit Theorem does not hold (see appendix).  In future research, we would also hope to better understand how redundancy, memory capacity, and predictive capacity vary with the number of neurons, as we ran into significant numerical integration difficulties here. Based on the work shown here, these changes would result in model time cells with nontrivial optimal time constants, as could be expected \cite{still2010optimal, marzen2016predictive, chalk2018toward}.

In conclusion, we have provided a quantitative framework for predicting optimal time constants of time cells that we hope will prove useful for those in neuroscience.

\acknowledgments

The authors thank MIT Physics of Living Systems for its hospitality during visits and Marc Howard for enlightening conversations.  A. H. and S. E. M. were funded by the Air Force
Office of Scientific Research under award number FA9550-19-1-0411.

\bibliography{optimaltimecells}

\appendix
\onecolumngrid

\section{Derivation of MC and PC in continuous time}

Say we have $m$ time cells, where the $i^{th}$ time cell's activity is given in the main text:
\begin{equation}
f_i(t) = \int_{-\infty}^{t} x(t') e^{-s_i(t-t')} dt'.
\end{equation}
In this section, we calculate closed-form expressions for $MC$ and $PC$, which were defined in the main text as well.  Throughout, we assume stationarity, and we assume that the input's mean value is $0$ and that its variance is $1$.

Recall that, in this case, the memory function is
\begin{equation}
m(\tau) = p_{\tau}^{\top} C^{-1} p_{\tau}
\end{equation}
where
\begin{equation}
p_{\tau} = \langle x(t-\tau) f(t) \rangle_t,~C = \langle f(t) f(t)^{\top} \rangle_t.
\end{equation}
We integrate this memory function from $\tau$ being $\infty$ to $0$ to get $MC$, and from $0$ to $\infty$ to get $PC$.  Using our earlier expression for the activity $f$, we find that
\begin{eqnarray}
(p_{\tau})_i &=& \langle x(t-\tau) \int_{-\infty}^t x(t') e^{-s_i(t-t')} dt' \rangle_t \\
&=& \int_{-\infty}^t \langle x(t-\tau) x(t')\rangle e^{-s_i(t-t')} dt' \\
&=& \int_{-\infty}^t R(t-\tau-t') e^{-s_i(t-t')} dt' \\
&=& \int_0^{\infty} R(t'-\tau) e^{-s_i t'} dt'
\end{eqnarray}
and
\begin{eqnarray}
C_{ij} &=& \langle f_i(t) f_j(t) \rangle_t \\
&=& \langle \int_{-\infty}^t x(t') e^{-s_i(t-t')} dt' \int_{-\infty}^t x(t'') e^{-s_j(t-t'')} dt'' \rangle_t \\
&=& \int_{-\infty}^t \int_{-\infty}^t e^{-s_i(t-t')} e^{-s_j(t-t'')} R(t'-t'') dt' dt'' \\
&=& \int_0^{\infty} \int_0^{\infty} e^{-s_i t'} e^{-s_j t''} R(t'-t'') dt' dt''.
\end{eqnarray}
At this point, we recall that
\begin{equation}
R(t) = \int_0^{\infty} F(\lambda) e^{-\lambda|t|} d\lambda.
\end{equation}
(One can also derive similar expressions by using the Fourier transform.)
Plugging this in, we have
\begin{eqnarray}
C_{ij} &=& \int_0^{\infty} \int_0^{\infty} e^{-s_i t'} e^{-s_j t''} \int_0^{\infty} F(\lambda) e^{-\lambda|t'-t''|} d\lambda dt' dt'' \\
&=& \int_0^{\infty} \int_0^{\infty} \int_0^{\infty} F(\lambda) e^{-s_i t'} e^{-s_j t''}  e^{-\lambda |t'-t''|} dt' dt'' d\lambda \\
&=& \int_0^{\infty} F(\lambda) \frac{2\lambda +s_i+s_j}{(s+s_i)(s+s_j)(s_i+s_j)} d\lambda.
\end{eqnarray}
When $\tau>0$, we find that
\begin{eqnarray}
(p_{\tau})_i &=& \int_0^{\infty} \int_0^{\infty} F(\lambda) e^{-\lambda |t'-\tau|} d\lambda e^{-s_i t'} dt' \\
&=& \int_0^{\infty} F(\lambda) \left(\int_0^{\tau} e^{-s_it'} e^{-\lambda(\tau-t')} dt' + \int_{\tau}^{\infty} e^{-s_it'} e^{-\lambda(t'-\tau)} dt'\right) d\lambda\\
&=& \int_0^{\infty} F(\lambda) \frac{2\lambda e^{-s_i\tau}-(\lambda+s_i)e^{-\lambda \tau}}{\lambda ^2-s_i^2} d\lambda.
\end{eqnarray}
Otherwise, we find that
\begin{eqnarray}
(p_{\tau})_i &=& \int_0^{\infty} \int_0^{\infty} F(\lambda) e^{-\lambda|t'-\tau|} d\lambda e^{-s_i t'} dt' \\
&=& \int_0^{\infty} F(\lambda) \frac{e^{-\lambda \tau}}{\lambda+s_i} d\lambda.
\end{eqnarray}
Using our formula for the memory function $m(\tau)$ and for $MC$, $PC$, we have
\begin{eqnarray}
MC &=& \int_0^{\infty} m(\tau) d\tau \\
&=& \int_0^{\infty}  \sum_{i,j} (p_{\tau})_i (C^{-1})_{ij} (p_{\tau})_j d\tau \\
&=& \sum_{i,j} (C^{-1})_{ij} \int_0^{\infty} \left( \int_0^{\infty} F(\lambda) \frac{2\lambda e^{-s_i\tau}-(\lambda+s_i)e^{-\lambda \tau}}{\lambda ^2-s_i^2} d\lambda\right) \left( \int_0^{\infty} F(\lambda) \frac{2\lambda'e^{-s_j\tau}-(\lambda'+s_j)e^{-\lambda'\tau}}{(\lambda')^2-s_j^2} d\lambda'\right) d\tau \nonumber \\
&=& \int_{\lambda=0}^{\infty}\int_{\lambda'=0}^{\infty}\left(\frac{F(\lambda)F(\lambda')}{\left(\lambda^{2}-s_{i}^{2}\right)\left(\lambda'^{2}-s_{j}^{2}\right)}\left[\frac{4\lambda\lambda'}{s_{i}+s_{j}}-\frac{2\lambda(\lambda'+s_{j})}{s_{i}+\lambda'}-\frac{2\lambda'(\lambda+s_{i})}{\lambda+s_{j}}+\frac{(\lambda+s_{i})(\lambda'+s_{j})}{\lambda+\lambda'}\right]\right)d\lambda d\lambda'\\
&=& 1^{\top} \left( C^{-1} \odot D_{MC} \right) 1
\end{eqnarray}
with $D_{MC}$ having entries given in the main text, Eq. \ref{eq:DMC}.
Similarly,
\begin{eqnarray}
PC &=& \int_{-\infty}^0 m(\tau) d\tau \\
&=& \sum_{i,j} (C^{-1})_{ij}\int_{-\infty}^0\left( \int_0^{\infty} F(\lambda) \frac{e^{-\lambda \tau}}{\lambda+s_i} d\lambda\right) \left( \int_0^{\infty} F(\lambda') \frac{e^{-\lambda\tau}}{\lambda'+s_i} d\lambda'\right) d\tau \\
&=& \sum_{i,j} (C^{-1})_{ij} \int_0^{\infty} \int_0^{\infty}\frac{F(\lambda) F(\lambda')}{(\lambda+\lambda')(\lambda+s_i)(\lambda'+s_j)} d\lambda d\lambda' \\
&=& 1^{\top} \left(C^{-1} \odot D_{PC}\right) 1,
\end{eqnarray}
with $D_{PC}$ given in Eq. \ref{eq:DPC} of the main text.

\section{Extension to the case of multi-dimensional input}

Much of the sensory input that we receive, e.g. natural video, is high-dimensional.  To that end, we consider extending our analysis to the case of high-dimensional inputs, such that neuron $i$ has activity $f_i$ given by
\begin{equation}
    \frac{df_i}{dt} = -s_i f_i + v_i^{\top} x.
\end{equation}
Now, $v_i$ is a vector such that the input $x$, also a vector, is converted into a scalar.  In this way, it is relatively straightforward to alter the model of time cells.

However, the definitions of memory and predictive capacity need to be altered accordingly.  We consider trying to predict $x_j(t+\tau)$ from $\vec{f}(t)$ and to remember $x_j(t-\tau)$ from $\vec{f}(t)$, and calculating $PC_j$ and $MC_j$, respectively, by integrating the squared correlation coefficient over all $\tau$.  We then sum $MC_j$ and $PC_j$ over all dimensions $j$ in order to get a final $MC$ and $PC$.

Let $(p_{\tau})_{i,j} = \langle x_j(t+\tau) f_i(t)\rangle_t$ and $C_{i,j} = \langle f_i(t) f_j(t) \rangle_t$, the latter as before, but the former with the additional index corresponding to the dimension of the input. Also, let $R_{j,k}(\tau)=\langle x_{j}(t)x_{k}(t-\tau) \rangle$ and $\overleftrightarrow{R}(\tau)$ be the matrix valued autocorrelation function with $R_{j,k}(\tau)$ as the entries.  Some algebra similar to that of the appendix above and not shown here gives
\begin{eqnarray}
(p_{\tau})_{i,j} = \int_0^{\infty} e^{-s_i t'}  \left[\overleftrightarrow{R}(t'-\tau)\vec{v_i}\right]_j dt'.
\end{eqnarray}
Note that $\left[\overleftrightarrow{R}(t'-\tau)\vec{v_i}\right]_j$ denotes the $j$'th entry of the enclosed matrix-vector product. More straightforward algebra similar to that of the one dimensional case in the previous appendix gives
\begin{eqnarray}
C_{i,j} &=& \int_0^{\infty} \int_0^{\infty} e^{-s_i t'} e^{-s_j t''} v_i^{\top} \overleftrightarrow{R}(t'-t'') v_j dt' dt'',
\end{eqnarray}
 Together, these determine the memory function for the $j^{th}$ element of the input:
\begin{equation}
    m_j(\tau) = (\vec{p}_{\tau}^{j})^{\top} C^{-1} \vec{p}_{\tau}^{j}
\end{equation}
and from there, the total memory capacity and predictive capacity:
\begin{equation}
    MC = \sum_j \int_{-\infty}^0 m_j(\tau) d\tau,~PC = \sum_j \int_0^{\infty} m_j(\tau) d\tau.
\end{equation}
In other words, we can understand the effect of spatial correlations on memory and predictive capacity by understanding its effects on $p_{\tau}$ and $C$.

We have just shown that only the (spatio-temporal) autocorrelation function is relevant for these metrics.  And, furthermore, if the temporal component is constant or nearly constant across dimensions of the input, we will find that $\overleftarrow{R}(\tau) = S g(\tau)$, where $S$ is the spatial covariance matrix and $g(\tau)$ represents the temporal component of autocorrelation function.  In such a case, under some conditions specified below, the analysis of optimal forgetting rates will not be governed by spatial patterns, but by $g(\tau)$.  For instance, we find
\begin{eqnarray}
(p_{\tau})_{i,j} &=& \left( S v_i \right)_j\cdot\int_0^{\infty} e^{-s_i t} g(t-\tau)dt  ,
\end{eqnarray}
so that $p_{\tau}$'s $\tau$ dependence is strongly governed by $g(t)$, and
\begin{eqnarray}
C_{i,j} &=& \int_0^{\infty} \int_0^{\infty} e^{-s_i t'} e^{-s_j t''}  g(t'-t'') v_i^{\top} S v_j dt' dt''.
\end{eqnarray}
Note that this splits into an element-wise product of a spatial component (with elements $v_i^{\top} S v_j$) and a temporal component (with elements $\int_0^{\infty} \int_0^{\infty} e^{-s_i t'} e^{-s_j t''}  g(t'-t'') dt' dt''$).
Thus, when $\overleftarrow{R}(\tau)$ admits (or approximately admits) such a decomposition, we find that $C$ and $~(p_{\tau})_j$ is roughly the same as that for a single pixel, and so analysis of one pixel is equivalent to an analysis of all pixels.  Natural video may fall into this class of inputs after spatial processing by the visual cortex if receptive fields are sufficiently diffuse.  More research will need to be conducted to elucidate the effects of the spatial component on maximization of $MC$ or $PC$.

\section{Optimality of the Laplace transform}

In this section, we consider a slightly more general model for how neuronal activity evolves:
\begin{equation}
\frac{df_i}{dt} = -\omega_i f_i + \sum_j J_{ij} \phi(f_j) + x(t).
\end{equation}
Due to the nonlinearity $\phi$, the neuronal activities will no longer be Laplace transforms of the input.

This is intractable unless we make some assumptions.  As such, we assume that there are a very large number of neurons $N$, and that $J_{ij}$ connections are randomly chosen from some distribution, where the mean is $0$ and the variance is $\sigma_J^2/N$.  Then, $\eta_i (t) = \sum_j J_{ij}\phi(x_j)$ is normally distributed according to the Central Limit Theorem.  If this is the case, then the nonlinear term in the new evolution equation corresponds to Gaussian noise, and the now-linear system with Gaussian noise can still be analyzed.

We follow Ref. \cite{Rajan} in our treatment.  We first characterize the noise properties:
\begin{eqnarray}
\langle \eta_i(t) \rangle &=& \langle \sum_j J_{ij} \phi(x_j) \rangle = 0
\end{eqnarray}
and-- assuming that $N$ is so large that $J_{ij}$ is roughly uncorrelated with $\phi(f_j),~\phi(f_i)$-- we find
\begin{eqnarray}
\langle \eta_i(t) \eta_j(t+\tau) \rangle &=& \left\langle \left( \sum_k J_{ik} \phi(x_k)\right) \left( \sum_{k'} J_{ik'} \phi(x_{k'})\right) \right\rangle \\
&=& \sum_{k,k'}\langle J_{ik} J_{jk'} \rangle \langle \phi(x_k) \phi(x_{k'})\rangle \\
&=& \sum_{k,k'} \frac{1}{N}(\delta_{i,j} \delta_{k,k'} \sigma_J^2) \langle \phi(x_k) \phi(x_{k'}) \rangle
\\
&=& \frac{\delta_{i,j} }{N} \sum_k \sigma_J^2 \langle \phi(x_k(t)) \phi(x_k(t+\tau)) \rangle \\
&=& \delta_{i,j} \sigma_J^2 \langle \phi(x(t))\phi(x(t+\tau)) \rangle
\end{eqnarray}
Since the nonlinear term corresponds to Gaussian noise and the $\omega_i f_i$ term is linear, then given the input, $f_i$ is normally distributed with mean $0$ (since $\langle x\rangle = \langle \eta\rangle= 0$) and a covariance between $f_i(t)$ and $f_i(t+\tau)$ of $C(\tau)$:
\begin{equation}
    C(\tau) := \langle f_i(t+\tau) f_i(t) \rangle.
\end{equation}
We then define
\begin{equation}
K(\tau) : = \langle \phi(x(t))\phi(x(t+\tau)) \rangle 
\end{equation}
which becomes
\begin{equation}
    K(\tau)= \int \int \phi(x) \phi(y) \frac{\exp\left(\frac{1}{2} \frac{C(0) x^2 - 2 C(\tau) xy + C(0) y^2}{C(0)^2-C(\tau)^2}\right)}{2\pi\sqrt{|C(0)^2-C(\tau)^2|}} dx dy.
\end{equation}
If we can now find a relationship between $C(\tau)$ and $K(\tau)$, we will be able to solve for both.  To do this, we return to the original evolution equation and solve explicitly for $x_i(t)$:
\begin{eqnarray}
\dot{x_i} +\omega_i x_i &=& \eta_i(t) + f(t) \\
e^{-\omega_i t} \frac{d}{dt} \left( e^{\omega_i t} x_i\right) &=& \eta_i + f \\
\frac{d}{dt} \left( e^{\omega_i t} x_i \right) &=& e^{\omega_i t} \left( \eta_i + f\right) \\
e^{\omega_i t} x_i(t) &=& \int_{-\infty}^t e^{\omega_i s} \left(\eta_i(s) + f(s) \right) ds \\
x_i(t) &=& \int_{-\infty}^t e^{-\omega_i (t-s)} \left(\eta_i(s) + f(s) \right) ds.
\end{eqnarray}
We know that $x_i$ given $f$ is normally distributed.  Its mean is clearly $0$.  $C(\tau)$ is straightforwardly obtained:
\begin{eqnarray}
\langle x_i(t) x_i(t+\tau) \rangle &=& \left\langle \left(\int_{-\infty}^t e^{-\omega_i (t-s)} \left(\eta_i(s) + f(s) \right) ds\right) \left(\int_{-\infty}^{t+\tau} e^{-\omega_i (t+\tau-s)} \left(\eta_i(s) + f(s) \right) ds\right) \right\rangle \\
&=& \int_{-\infty}^{t} \int_{-\infty}^{t+\tau} e^{-\omega_i(t-s)} e^{-\omega_i(t+\tau-s')} \langle (\eta_i(s) + f(s)) (\eta_i(s') + f(s'))\rangle ds ds' \\
C(\tau) &=& \int_{-\infty}^{t} \int_{-\infty}^{t+\tau} e^{-\omega_i(t-s)} e^{-\omega_i(t+\tau-s')} \left(\sigma_J^2 K(s-s') + R(s-s')\right) ds ds'  \\
&=& \int_{-\infty}^0 \int_{-\infty}^{\tau} e^{\omega_i s} e^{-\omega_i (\tau-s')} \left(\sigma_J^2 K(s-s') + R(s-s') \right) ds ds',
\end{eqnarray}
where $R$ is the autocorrelation function of the input.
In order to calculate $PC$ and $MC$, we need
\begin{eqnarray}
p_{\tau} &=& \langle x(t) f_i(t+\tau) \rangle \\
&=& \left\langle x(t)  \int_{-\infty}^{t+\tau} e^{-\omega_i (t+\tau-s)} \left(\eta_i(s) + x(s) \right) ds \right\rangle \\
&=& \int_{-\infty}^{t+\tau} e^{-\omega_i (t+\tau-s)} \langle x(t) \eta_i(s)\rangle + \langle x(t) x(s) \rangle ds \\
&=& \int_{-\infty}^{\tau} e^{\omega_i (s-\tau)} R(s) ds.
\end{eqnarray}
When $i\neq j$, we find that the activities of the two neurons are related via
\begin{eqnarray}
\langle f_i(t) f_j(t+\tau) \rangle &=& \left\langle \left(\int_{-\infty}^t e^{-\omega_i (t-s)} \left(\eta_i(s) + x(s) \right) ds\right) \left(\int_{-\infty}^{t+\tau} e^{-\omega_j (t+\tau-s')} \left(\eta_j(s') + x(s') \right) ds\right) \right\rangle \\
&=& \int_{-\infty}^t \int_{-\infty}^{t+\tau} e^{-\omega_i(t-s)} e^{-\omega_j (t+\tau-s')} \left(\langle \eta_i(s) \eta_j(s')\rangle + \langle \eta_i(s) x(s') \rangle + \langle \eta_j(s) x(s') \rangle + R(s-s')\right) ds ds' \nonumber \\
&=& \int_{-\infty}^t \int_{-\infty}^{t+\tau} e^{-\omega_i (t-s)} e^{-\omega_j (t+\tau-s')} R(s-s') ds ds' \\
&=& \int_{-\infty}^0 \int_{-\infty}^{\tau} e^{\omega_i s} e^{-\omega_j (\tau-s')} R(s-s') ds ds'.
\end{eqnarray}
This gives us our second relationship between $C(\tau)$ and $K(\tau)$.

Thus, we have
\begin{eqnarray}
C(\tau) &=& \int_{-\infty}^0 \int_{-\infty}^{\tau} e^{\omega_i s} e^{-\omega_i (\tau-s')} \left(R(s-s') + \sigma_J^2 K(s-s')\right) ds ds' \\
K(\tau) &=& \int\int \phi(x) \phi(y) \frac{\exp(-\frac{1}{2} \frac{C(0) x^2 - 2C(\tau) xy + C(0) y^2}{C(0)^2-C(\tau)^2})}{2\pi\sqrt{|C(0)^2-C(\tau)^2|}} dx dy
\end{eqnarray}
as the self-consistent equations.

In principle, that does it, but we seek some understanding from this math.  To simplify things, we now assume that all the neurons have the same timescale $\omega$, giving
\begin{equation}
(\vec{p}_{\tau})_i = \int_{-\tau}^{\infty} e^{-\omega (s+\tau)} R(s) ds
\end{equation}
and
\begin{eqnarray}
(Cov)_{ij} &=& \langle f_i(t) f_j(t) \rangle \\
&=& \int_{-\infty}^0 \int_{-\infty}^0 e^{\omega_i s+\omega_j s'} (\sigma_J^2 K(s-s') \delta_{ij} + R(s-s')) ds ds' \\
&=& \int_0^{\infty} \int_0^{\infty} e^{-\omega (s + s')} R(s-s') ds ds' + \sigma_J^2 \delta_{i,j} \int_0^{\infty} \int_0^{\infty} e^{-\omega (s+s')} K(s-s') ds ds',
\end{eqnarray}
which is the covariance matrix for $f(t)$.
Meanwhile, we still have
\begin{eqnarray}
C(\tau) &=& \int_{-\infty}^0 \int_{-\infty}^{\tau} e^{\omega (s+s'-\tau)} \left(R(s-s') + \sigma_J^2 K(s-s')\right) ds ds' \\
K(\tau) &=& \int\int \phi(x) \phi(y) \frac{\exp(-\frac{1}{2} \frac{C(0) x^2 - 2C(\tau) xy + C(0) y^2}{C(0)^2-C(\tau)^2})}{2\pi\sqrt{|C(0)^2-C(\tau)^2|}} dx dy
\end{eqnarray}
as the self-consistent equations.

Notice that
\begin{equation}
Cov = R_0 1_N + \sigma_J^2 K_0 I_N
\end{equation}
where $1_N$ is a $N\times N$ matrix of all $1$'s, and $I_N$ is the $N\times N$ identity matrix.  We also have
\begin{equation}
\vec{p}_{\tau} = R_{\tau} 1_N
\end{equation}
where now $1_N$ is the length $N$ vector of all $1$'s.  Then we have
\begin{eqnarray}
PC_{\tau} &=& R_{\tau}^2 1_N^{\top} \left(R_0 1_N + \sigma_J^2 K_0 I_N\right)^{-1} 1_N \\
&=& R_{\tau}^2 1_N^{\top} \left( \sigma_J^{-2} K_0^{-1} (I_N + \frac{R_0}{\sigma_J^2 K_0} 1_N)^{-1} \right) 1_N \\
&=& \frac{R_{\tau}^2}{\sigma_J^2 K_0} 1_N^{\top} \left(\sum_{k=0}^{\infty} (-1)^k \left(\frac{R_0}{\sigma_J^2 K_0} 1_N\right)^k\right) 1_N \\
&=& \frac{R_{\tau}^2}{\sigma_J^2 K_0} 1_N^{\top} \left(\sum_{k=0}^{\infty} (-\frac{R_0 N}{\sigma_J^2 K_0})^k 1_N \right) 1_N \\
&=& \frac{R_{\tau}^2 N^2}{\sigma_J^2 K_0} \left( \sum_{k=0}^{\infty} \left(-\frac{R_0 N}{\sigma_J^2 K_0}\right)^k\right) = \frac{R_{\tau}^2 N^2}{\sigma_J^2 K_0} \left(1+\frac{R_0 N}{\sigma_J^2 K_0}\right)^{-1}
\end{eqnarray}
where
\begin{eqnarray}
R_{\tau} &=& \int_{0}^{\infty} e^{-\omega s} R(s-\tau) ds \\
R_0 &=& \int_0^{\infty} \int_0^{\infty} e^{-\omega (s+s')} R(s-s') dsds' \\
K_0 &=& \int_0^{\infty} \int_0^{\infty} e^{-\omega (s+s')} K(s-s') ds ds'.
\end{eqnarray}
Clearly, we can increase $N$ arbitrarily and arbitrarily increase $m(\tau)$.  To get total $PC$, we integrate over $\tau$ and find the following:
\begin{eqnarray}
PC &=& \frac{N^2}{\sigma_J^2 K_0} \left(1+\frac{R_0 N}{\sigma_J^2 K_0}\right)^{-1} \int_0^{\infty} R_{\tau}^2 d\tau \\
&=& \frac{N^2}{\sigma_J^2 K_0} \left(1+\frac{R_0 N}{\sigma_J^2 K_0}\right)^{-1} \int_0^{\infty} \int_0^{\infty} \int_0^{\infty} e^{-\omega (s+s')} R(s-\tau) R(s'-\tau) ds ds' d\tau
\end{eqnarray}
If we look at how to maximize this, we see that there is a critical parameter
\begin{equation}
\rho = N/\sigma_J^2 K_0
\end{equation}
which gives
\begin{equation}
PC = \frac{N\rho}{1+R_0\rho} \int_0^{\infty} \int_0^{\infty} \int_0^{\infty} e^{-\omega (s+s')} R(s-\tau) R(s'-\tau) ds ds' d\tau.
\end{equation}
$PC$ is clearly maximized when $\rho\rightarrow \infty$, which can be achieved by: the number of nodes $N$ going to infinity, the nonlinearity weight variances $\sigma_J^2 \rightarrow 0$, or the nonlinearity-controlled $K_0 \rightarrow 0$.  When we are in that limit, we find
\begin{equation}
PC_{max} = N \frac{\int_0^{\infty} \int_0^{\infty} \int_0^{\infty} e^{-\omega (s+s')} R(s-\tau) R(s'-\tau) ds ds' d\tau}{\int_0^{\infty} \int_0^{\infty} e^{-\omega (s+s')} R(s-s') ds ds'}
\end{equation} 
Given that $\sigma_J \rightarrow 0$ is optimal, to maximize $PC$ in this admittedly limited setup, we should opt to minimize nonlinearities.

\section{Results for a different model time cell}

In this appendix, the activity of time cell $f(s)$ at time $t$ is a direct readout of $x(t-s)$.  Assuming continuity of $x(t)$, this is equivalent to assuming that the activity of time cell $f(s)$ at time $t$ is a direct readout of $\frac{1}{s_1+s_2}\int_{t-s_1}^{t-s_2} x(s') ds'$ for some $s$ by the Intermediate Value Theorem.  When we refer to this neuron's timescale, we mean the delay time $s$.

If stationarity holds, then
\begin{eqnarray}
I[f(s_i);f(s_{i+1})] &=& I[x(t-s_i);x(t-s_{i+1})] \\
&=& I[x(0);x(s_{i+1}-s_i)].
\end{eqnarray}
Hence, equalizing redundancy between two neighboring neurons implies keeping $s_{i+1}-s_i$ a constant.  This is emphatically \emph{not} the logarithmic scaling of Ref. \cite{}.  One can extend this argument to any measure of redundancy, as any measure of redundancy as described in Ref. \cite{} is a function of the joint probability distribution $P(f(s_i),f(s_{i+1}))$ and hence subject to the restrictions of stationarity.  Note also that for almost all processes, $I[f(s_i);f(s_{i+1})]$ will tend to $0$ as $s_{i+1}-s_i$ increases to infinity.

The efficient coding hypothesis argument in the main text applies equally well, and in some ways more rigorously, to these model time cells.  Hence, redundancy reduction and efficient coding are equivalent for these model time cells as well.

In order to remember the entire past as well as possible, one would want to place the receptive fields of neurons as far back as possible, assuming that remembering what happened $3$ years ago was as important as remembering what happened $1$ day ago.  We would therefore expect that optimally, $s_i\rightarrow\infty$.

Finally, for most signals, the recently observed signal is a better clue to the future than a previously observed signal, as discussed in the main text.  We'd therefore expect $s_i\rightarrow 0$ optimally.

\section{Additional analysis of memory capacity}

In this appendix, we analyze memory capacity of optimal time cells for a few different types of input statistics.  In all situations, we consider the one-node (one neuron, one time cell) case.  For all these input types, we find that $MC$ is maximized as $s\rightarrow 0$.

Let's start with the simplest possible input: a Markovian signal with timescale $\lambda_0$:
\[
R(t)=e^{-\lambda_{0}|t|}.
\]
In this case,
\[
MC(s)=\frac{4\lambda_{0}+s}{2\lambda_{0}^{2}+2\lambda_{0}s}.
\]
The derivation of this is somewhat tricky, but achievable by using $F(\lambda)=\delta(\lambda-\lambda_0)$ and carefully keeping track of singularities.  One can check that there is a maximum of $MC$ as $s\rightarrow 0$.

When $R(t)$ is instead a mixture of timescales 
\[
R(t)=\frac{1}{2}e^{-\lambda_{0}|t|}+\frac{1}{2}e^{-\lambda_{1}|t|}
\]
then
\begin{eqnarray*}
MC(s)&=&\frac{4\lambda_{0}\lambda_{1}\left(\lambda_{0}+\lambda_{1}\right)^{3}+\left(\lambda_{0}^{4}+17\lambda_{0}^{3}\lambda_{1}+36\lambda_{0}^{2}\lambda_{1}^{2}+17\lambda_{0}\lambda_{1}^{3}+\lambda_{1}^{4}\right)s+2\left(\lambda_{0}+\lambda_{1}\right)\left(\lambda_{0}^{2}+10\lambda_{0}\lambda_{1}+\lambda_{1}^{2}\right)s^{2}}{4\lambda_{0}\lambda_{1}\left(\lambda_{0}+\lambda_{1}\right)\left(\lambda_{0}+s\right)\left(\lambda_{1}+s\right)\left(\lambda_{0}+\lambda_{1}+2s\right)} \nonumber \\
&& + \frac{\left(\lambda_{0}^{2}+6\lambda_{0}\lambda_{1}+\lambda_{1}^{2}\right)s^{3}}{4\lambda_{0}\lambda_{1}\left(\lambda_{0}+\lambda_{1}\right)\left(\lambda_{0}+s\right)\left(\lambda_{1}+s\right)\left(\lambda_{0}+\lambda_{1}+2s\right)}
\end{eqnarray*}
And for the case that $F(\lambda)=\begin{cases}
\frac{1}{b} & 0<x<b\\
0 & otherwise
\end{cases}$, we find that memory is maximized as $s\to0,$
$MC\to\infty$. This is a special case of the class of $F(\lambda)$
which takes on the form 
\[
F(\lambda)=\begin{cases}
\frac{1}{b-a} & a<x<b\\
0 & otherwise
\end{cases}
\]

which produces autocorrelation functions of the form 
\[
R(t)=\frac{2e^{-(a+b)|t|}\left(e^{a|t|}\left(1+b|t|\right)-e^{b|t|}\left(1+a|t|\right)\right)}{\left(a^{2}-b^{2}\right)t^{2}}
\]

These correspond to autocorrelation functions produced by averaging
over an interval of characteristic timescales. In this case, $\lim_{s\to0^{+}}MC$
is available in closed form: 

\[
\lim_{s\to0^{+}}MC=\frac{2\log\left(\frac{a}{b}\right)}{a-b}.
\]

Setting $a\to0$ shows the logarithmic divergence of $MC$.

For an argument as to why $MC$ is optimized by sending $s\to0$
in general, consider the one-node case. Then $D_{MC}$ reduces to
\[
D_{MC}=\int_{0}^{\infty}\int_{0}^{\infty}F(\lambda)F(\lambda')\frac{2\lambda+2\lambda'+s}{s(\lambda+\lambda')(\lambda+s)(\lambda'+s)}d\lambda d\lambda'
\]

and has the series expansion centered at $s=0$ 
\[
\int_{0}^{\infty}\int_{0}^{\infty}\left[F(\lambda)F(\lambda')\left(\frac{2}{\lambda_{1}\lambda_{2}s}-\frac{2\lambda_{1}^{2}+3\lambda_{1}\lambda_{2}+2\lambda_{2}^{2}}{\lambda_{1}^{2}\lambda_{2}^{2}\left(\lambda_{1}+\lambda_{2}\right)}+O(s)\right)\right]d\lambda d\lambda'
\]

Multiplying by $C(s)^{-1}$, we therefore have that 
\[
MC=\int_{0}^{\infty}\int_{0}^{\infty}\left[\frac{F(\lambda)F(\lambda')}{C(s)}\left(\frac{2}{\lambda_{1}\lambda_{2}s}-\frac{2\lambda_{1}^{2}+3\lambda_{1}\lambda_{2}+2\lambda_{2}^{2}}{\lambda_{1}^{2}\lambda_{2}^{2}\left(\lambda_{1}+\lambda_{2}\right)}+O(s)\right)\right]d\lambda d\lambda'
\]

The two terms which, in $s$, have the largest contributions $\left(\frac{2}{\lambda_{1}\lambda_{2}s}-\frac{2\lambda_{1}^{2}+3\lambda_{1}\lambda_{2}+2\lambda_{2}^{2}}{\lambda_{1}^{2}\lambda_{2}^{2}\left(\lambda_{1}+\lambda_{2}\right)}\right)$,
are both maximized by setting $s\to0$. It is clear that for $\lambda_{1},\lambda_{2}>0$,
$\frac{2}{\lambda_{1}\lambda_{2}s}$ increases unboundedly by decreasing
$s\to0$. The constant coefficient in this expansion $-\frac{2\lambda_{1}^{2}+3\lambda_{1}\lambda_{2}+2\lambda_{2}^{2}}{\lambda_{1}^{2}\lambda_{2}^{2}\left(\lambda_{1}+\lambda_{2}\right)}$
is negative. Clearly, $\lambda_{1}^{2}\lambda_{2}^{2}\left(\lambda_{1}+\lambda_{2}\right)>0$,
and $2\lambda_{1}^{2}+3\lambda_{1}\lambda_{2}+2\lambda_{2}^{2}$ is
a positive definite quadratic form, making 
\[
\frac{2\lambda_{1}^{2}+3\lambda_{1}\lambda_{2}+2\lambda_{2}^{2}}{\lambda_{1}^{2}\lambda_{2}^{2}\left(\lambda_{1}+\lambda_{2}\right)}
\]
positive.  Hence, $MC$ is maximized as $s\rightarrow 0$.

\section{Maximizing predictive capacity}

\begin{figure}
\centering
\includegraphics[width=0.45\textwidth]{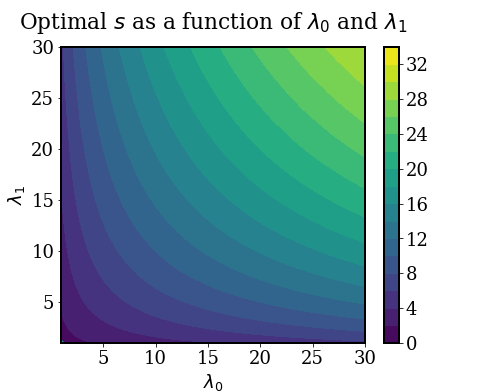}
\includegraphics[width=0.45\textwidth]{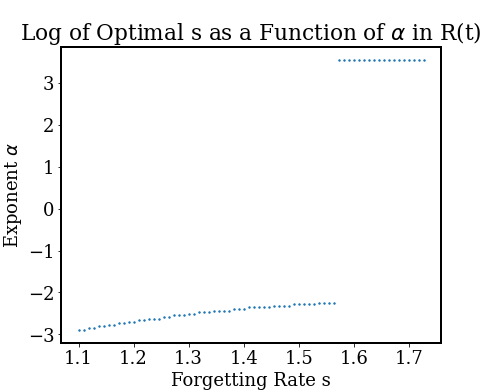}
\caption{(Left) A plot of the optimal forgetting rate $s$ for an environment with $R(t) = \frac{1}{2}\left(e^{-\lambda_1|t|}+e^{-|\lambda_2|t}\right)$, with $\lambda_1$ and $\lambda_2$ on the $x$-axis and $PC$ on the $y$-axis.  (Right) A plot of the log of the optimal forgetting rate, $\log s$, as a function of the parameter $\alpha$ for input autocorrelation functions of the form $R(t)=\frac{1}{1+|t|^\alpha}$.  Note the sudden increase at $\alpha=1.6$ from optimal $s$ being finite to optimal $s$ being the maximal $s$ we searched over.}
\label{fig:opts}
\end{figure}

In Fig. \ref{fig:opts}(left), we show the optimal forgetting rate of a model time cell when impinged upon by an input with autocorrelation function $R(t) = \frac{1}{2}\left(e^{-\lambda_1|t|}+e^{-|\lambda_2|t}\right)$.  This model time cell was augmented with another cell that stored the present value.

In Fig. \ref{fig:opts}(right), we show the optimal forgetting rate of a model time cell when impinged upon by an input with autocorrelation function $R(t)=\frac{1}{1+|t|^\alpha}$.  This model time cell was augmented with another cell that stored the present value.  Note that the optimal forgetting rate attains some intermediate, nontrivial value for most $\alpha$, indicating time-scale matching.  Furthermore, note that there appears to be a phase transition at $\alpha\approx 1.6$ at which point the model time cell desires to have a maximal forgetting rate.  The $\alpha$'s in our environment tend to be between $1$ and $2$ \cite{dong_atick_1995}.

\end{document}